# Influence of the carbon substitution on the critical current density and AC losses in $MgB_2$ single crystals

M Ciszek [1*], K Rogacki [1], K Oganisian [1], N D Zhigadlo [2] and J Karpinski [2]

[1] Polish Academy of Sciences, Institute of Low Temperature and Structure Research, ul. Okólna 2, 50-422 Wrocław, Poland

[2] Solid State Physics Laboratory, ETH 8093 Zürich, Switzerland

*) corresponding author, E-mail: M.Ciszek@int.pan.wroc.pl

**Abstract.** The DC magnetization and AC complex magnetic susceptibilities were measured for $MgB_2$ single crystals, unsubstituted and carbon substituted with the composition of $Mg(B_{0.94}C_{0.05})_2$. The measurements were performed in AC and DC magnetic fields oriented parallel to the c-axis of the crystals. From the DC magnetization loops and the AC susceptibility measurements, critical current densities ($J_c$) were derived as a function of temperature and the DC and AC magnetic fields. Results show that the substitution with carbon decreases $J_c$ at low magnetic fields, opposite to the well known effect of an increase of $J_c$ at higher fields. AC magnetic losses were derived from the AC susceptibility data as a function of amplitude and the DC bias magnetic field. The AC losses were determined for temperatures of 0.6 and 0.7 of the transition temperature $T_c$, so close to the boiling points of $LH_2$ and LNe, potential cooling media for magnesium diboride based composites. The results are analyzed and discussed in the context of the critical state model.

PACS: 74.25.Ha; 74.40.Ad; 74.25.Sv; 74.25.Nf

## 1 Introduction

The discovery of superconductivity in the magnesium diboride ($MgB_2$) compound [1] opened up new horizons for technical application of the material as well as for basic science and theoretical researches leading to explanation of the high temperature superconductivity phenomenon. From the practical application point of view the main merits of $MgB_2$ is relatively high critical temperature (about 40 K), absence of weak-links between grains, and low costs of the material. However it turned out that the pinning and the critical current densities of the superconductor are rather low. One of the very efficient method to improve the pinning is chemical doping and/or substitution of boron atoms by other atoms. Here, the most intensively studied substitution is carbon for boron which fills the $MgB_2$ hole-bands with electrons and also introduces scattering centers which may act in very different ways [2-6]. Substitution of carbon in place of boron increases considerably the upper critical field, $H_{c2}$, enhances the irreversibility field, $H_{irr}$, increases critical current densities at higher magnetic fields, lowers the $H_{c2}$ anisotropy, but simultaneously reduces $T_c$ [7-9]. There are many works addressing these issues, and for review see *e.g.* Refs. [10-11] and references therein. The other method to enhance pinning properties and thus the critical current density is proton or neutron irradiation, which introduces mainly atomic disorder in the $MgB_2$ structure [12-16].

For sintered bulk samples or in composite wires, the suppression of the critical current at grain boundaries is substantially less pronounced in $MgB_2$ than in other HTSC ceramics, *i.e.* there is no any drastic difference between inter- and intragrains critical current densities [17]. One can assume, that some results obtained for single crystals may be ascribed to large objects also, *e.g.* high quality bulk sintered samples or composite wires. Thus, energy loss measurements performed on single crystals allow to study in detail the intrinsic material properties and to verify this compound for applications with alternating currents. In the present work we focus on the critical current density and dissipation energy properties of $MgB_2$ single crystals at low magnetic fields. Both unsubstituted and carbon substituted crystals have been studied. The results are analyzed and discussed in the context of the critical state model for the ac susceptibility of samples with the rectangular shape. This issue remains practically unexplored with respect to $MgB_2$ single crystals at low magnetic fields.

## 2 Experimental

Objects of the research presented in the paper were two single crystals of magnesium diboride $MgB_2$, the first one as the chemically stechiometric unsubstituted compound (crystal MB) and the second one as substituted with carbon $Mg(B_{1-x}C_x)_2$ at x=0.06 (crystal MBC). More details of the synthesis process are published elsewhere [18]. The crystal structure was investigated on a single crystal X-ray diffractometer and the unit cell parameters were estimated. The carbon content in the MBC crystal was estimated from changes in the $a$ lattice parameter as described in our previous paper [8]. Some basic parameters of the crystals we have studied are given in Table 1. More physical properties describing superconductivity in these compounds are published elsewhere [8,9,13-16].

**Table 1**. Specification of $MgB_2$ (MB) and $Mg(B_{0.94}C_{0.06})_2$ (MBC) single crystals.

| Crystal | MB | MBC |
|---|---|---|
| Length, $2b$ (mm) | 0.805 | 0.529 |
| Width, $2a$ (mm) | 0.427 | 0.370 |
| Thickness, $2c$ (mm) | 0.067 | 0.069 |
| Volume (cm$^3$) | $23 \cdot 10^{-6}$ | $13.5 \cdot 10^{-6}$ |
| $T_c$ (onset) (K) | 38.5 | 33.4 |
| $J_{c0}$ (Am$^{-2}$) DC magn. | $1.0 \cdot 10^9$ (10 K) $7.9 \cdot 10^8$ (15 K) | $4.6 \cdot 10^8$ (10 K) |
| $J_{c0}$ (Am$^{-2}$) AC susc. | $1.0 \cdot 10^9$ (10 K) $8.6 \cdot 10^8$ (15 K) | $4.7 \cdot 10^8$ (10 K) |
| $\chi_0$ | 5.26 | 3.95 |

The single crystals used in our experiments were approximately plates in shape, with dimensions given in Table 1. The measurements of DC magnetization loops and AC susceptibility were performed with a commercial PPMS magnetometer at the temperature range from 5 K to 40 K, amplitudes of the AC magnetic field of 0.2-1.7 mT, DC bias magnetic fields up to 75 mT, and the frequencies range of 24-768 Hz. The field was oriented perpendicular to the $ab$ plane, which is the main plane of the $MgB_2$ single crystals. The critical temperatures ($T_c$) were determined from the onset of the AC magnetic susceptibility.

## 3 Results and discussion

### 3.1 DC magnetization measurements

Magnetic moment hysteresis loops were measured for each single crystals, at two temperatures, 10 and 15 K, in fields of up to 0.4 T. The magnetic field was applied at a sweep rate of 0.065 mT/sec, with a data sample rate of one point per 10 mT. The magnetization results are shown in Figs. 1(a) and 1(b), for the unsubstituted and the C-substituted $MgB_2$ single crystals, respectively. Estimated demagnetization factors from the magnetization curves in the Meissner state are; $N$=0.810 for the crystal MB and $N$=0.747 for the crystal MBC.

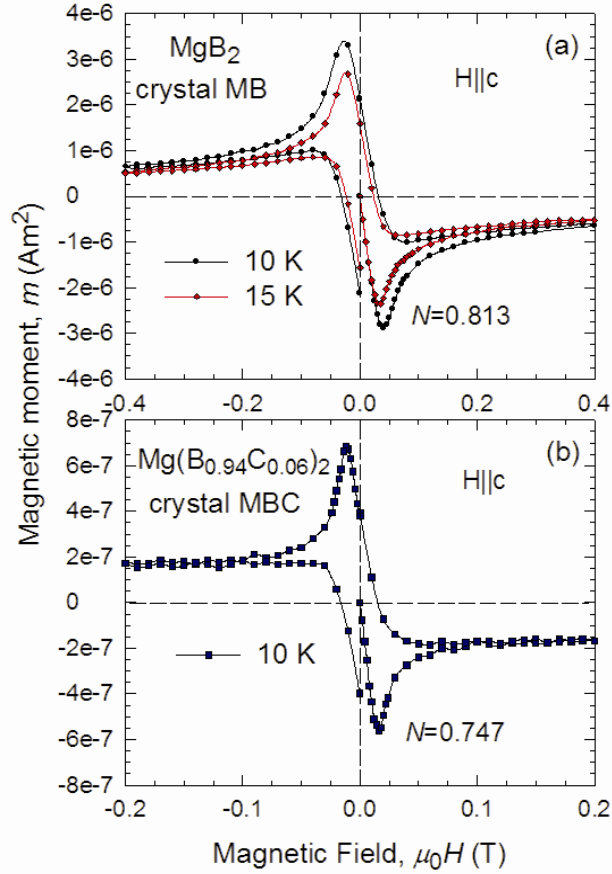

**Fig. 1**. Magnetic moment as a function of magnetic field for (a) the unsubstituted $MgB_2$ single crystal and (b) the C-substituted single crystal of composition $Mg(B_{0.94}C_{0.06})_2$. From the slope of the initial magnetization curve in the Meissner state demagnetization factors $N$ were calculated.

The magnetic hysteresis curves in Figure 1 indicate that the pinning force in the crystals is not very strong. For the crystal MB the irreversibility fields, determined from the magnetization loops are about 0.85 T and 0.55 T at temperatures 10 K and 15 K, respectively. For the crystal MBC the irreversible field is about 0.15 T at 10 K. Thus, for the low magnetic field region, as applied in our experiments, the carbon substitution decreases the irreversibility field significantly. Due to small dimensions of the single crystals, the criterion for the irreversible limit was taken as $\delta m = 5 \cdot 10^{-8}$ $Am^2$, which is much larger value than usually taken for polycrystalline samples with larger masses. Moreover, the measurements have been performed at low fields, *i.e.* before the fishtail effect is observed at much higher magnetic fields. Thus, the irreversibility fields we observe are much lower then those reported for higher fields [11,13-16].

The average critical current density of the square planar thick film samples can be calculated directly from the width of the hysteresis in magnetic moment $m$ of the sample, using formula derived on the base of the Bean critical state model. The measured hysteresis width, $\Delta m$, of the magnetic moment is related to the critical current density $J_c$ *via* equation:

$$J_c(B) = \frac{6|m^+(B) - m^-(B)|}{(3a-b)b^2 c} = \frac{6\Delta m(B)}{(3a-b)b^2 c} \quad (1)$$

where $a$ is the sample's width, $b$ its length ($b>a$) and $c$ is the sample's thickness. The negative and positive superscripts refer to the increasing and decreasing field regions of the hysteresis loop measurements, respectively [19]. Using equation (1), the critical current densities were derived from the magnetization curves (Fig. 1) and are depicted in Figure 2. The critical current densities calculated from the magnetic moment trapped in the remnant state ($B=0$), $J_{c0}$, are about $1.0 \cdot 10^9 A/m^2$ and $4.6 \cdot 10^8$ $A/m^2$ at 10K, for the unsubstituted and C-substituted $MgB_2$ single crystal, respectively (see Table 1). Generally, most of the experimental results show exponential or inverse-power dependence of $J_c$ on the external magnetic field [11]. For our results, we found that the $J_c(B)$ dependence can be described well by the Kim-like critical state model relation in the power form:

$$J_c(B) = J_{c0} \cdot \left[1 + (B/B_0)^\beta\right]^{-1} \qquad (2)$$

where $J_{c0}$ is a critical current density at zero magnetic field, and $B_0$ and $\beta$ are parameters depending on the microstructure properties of a superconductor. As showed by lines in Figure 2 the equation (2) fits reasonably well to our $J_c(B)$ experimental points. For the unsubstituted $MgB_2$ single crystal, the best fits were obtained for parameters $B_0$=43 mT and $\beta$=2.15 at 10 K, and $B_0$=36 mT and $\beta$=2 at 15 K. The parameter $B_0$, related to the pinning force, decreases with increasing temperature, as expected when no peak effect (fishtail for magnetization loops) is observed. For the C-substituted crystal, at temperature of 10K the best fit was found for $B_0$=21 mT and $\beta$=2.6. It is clearly seen, that the carbon doping process significantly reduces the critical current density at low magnetic fields. The observed change in the $\beta$ parameter from the initial value of 2.15 (unsubstituted crystal) to $\beta$=2.6 shows that $J_c$ of the C-substituted crystal drops slightly faster with an applied magnetic field.

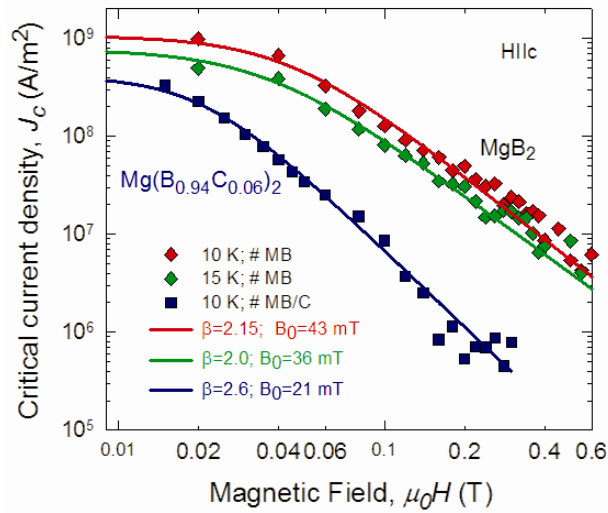

**Fig. 2**. Critical current density $J_c$ *versus* magnetic field $H$, for the unsubstituted $MgB_2$ single crystal at temperatures 10 and 15 K and for the C-substituted $Mg(B_{0.94}C_{0.06})_2$ single crystal at 10 K. The data were calculated from magnetization curves (Fig. 1). Solid lines are fits of equation (2) to the experimental data (for parameters see text).

**3.2 AC susceptibility measurements**

Dynamic magnetic properties of the single crystals of $MgB_2$ have been studied in terms of AC magnetic complex susceptibilities. The measurements were performed for the AC driving field amplitudes $b_0$, ranging from 0.2 mT to 1.7 mT, and frequencies of 24 Hz to 768 Hz. Additional DC bias field of the order up to 75 mT was applied. The AC and DC magnetic fields were oriented perpendicular to the *ab* plane of the samples. The temperature was changed with the rate of 0.1 K/min. Here we present only AC complex susceptibility data for the fundamental frequency of the exciting magnetic field.

Temperature dependencies of the volume magnetic AC susceptibility, $\chi = \chi' + i\chi''$, of the investigated single crystals are shown in Figure 3. The general behavior of the AC susceptibility is typical for type-II superconductors. The real part of the complex susceptibility $\chi'$ shows the dispersive magnetic response of a crystal and are related directly to the phase transition between normal and superconducting states. The imaginary part of the susceptibility, $\chi''$, which represents the critical current density and thus the hysteresis losses, exhibits only one clear maximum, what indicates the lack of weak link effects and a single phase quality of the crystals. The loss peak shifts to lower temperature when AC magnetic field amplitude increases. The data in Figure 3 are normalized by a constant $\chi_0$, so called external susceptibility, which represents demagnetization effects due to the surface screening current in the Meissner state [20]. For relatively simple crystal's geometry, the parameter $\chi_0$ is related to demagnetization factor $N$ as $\chi_0=-1/(1-N)$. In our experiments the values of $N$ were determined from the DC magnetization curves, as described in section 3.1. Precise determination of $\chi_0$ is very important as any error in its evaluation directly influences the values of $\chi'$ and $\chi''$ which are derived from the measurements.

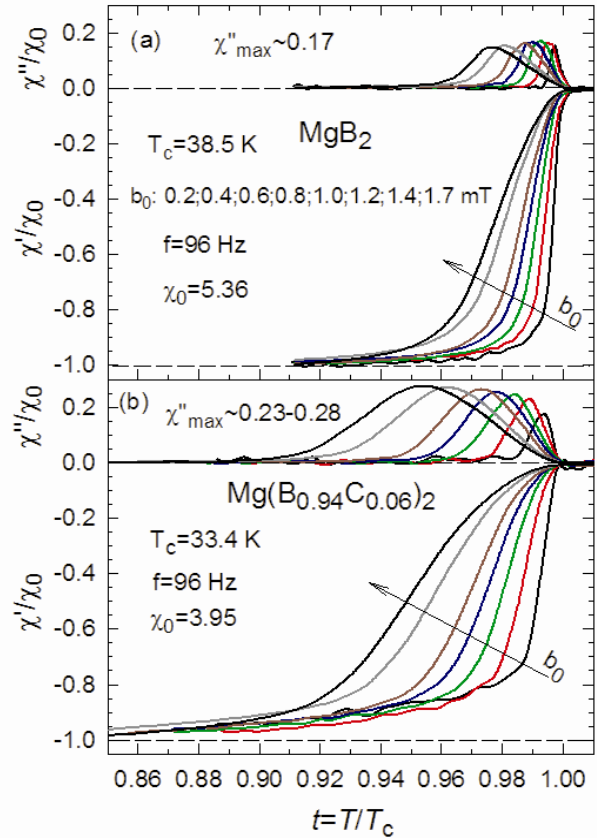

**Fig. 3**. Complex AC susceptibilities $\chi'$ and $\chi''$ as a function of temperature $t=T/T_c$ for (a) unsubstituted $MgB_2$ single crystal and (b) C-substituted $Mg(B_{0.94}C_{0.06})_2$ single crystal. The values of the susceptibilities are normalized to the external susceptibility parameter $\chi_0$, as described in text.

No significant frequency dependence of $\chi(t)$ was observed at the low frequencies, *i.e.* from 24 Hz to about 196 Hz, and at low amplitudes of the AC magnetic field, *i.e.* from 0.2 to about 1.0 mT. However, at higher amplitudes (above 1.0 mT) and higher frequencies (above 196 Hz) some slight frequency dependence is observed. Here we present and discuss only the results for low frequency range (from 24 Hz to 96 Hz), where pure hysteretic behavior of the AC susceptibility is observed.

In Figure 3 it is clearly visible that the C substitution in the $MgB_2$ crystal worsens its superconducting quality: *e.g.* lowers the critical transition temperature, and simultaneously broadens the transition temperature width. For a certain magnetic field amplitude, the $\chi''$ values form a single peak which maximum, $\chi''_m$, occurs at a certain temperature, $T_m$. The temperature $T_m$ shifts towards lower values when the amplitude of the field increases. Also the width of $\chi''$ maximum becomes wider as $b_0$ increases. This feature of broadening is much more pronounced in the MBC crystal. For example, from susceptibility data ($b_0$=0.2 mT) the transition width $\Delta T_c$ (10-90% criterion) is 0.45 K for the MB crystal, whereas for the crystal with carbon, this value accounts 1.25 K. As the shift of $T_m$ is a measure of the pinning force changes, the broadening of the transition temperature width and its lowering observed for the C-substituted crystal can be attributed to the introduction of disorders to the crystal structure, similar to disorders introduced by proton or neutron irradiations [12-16].

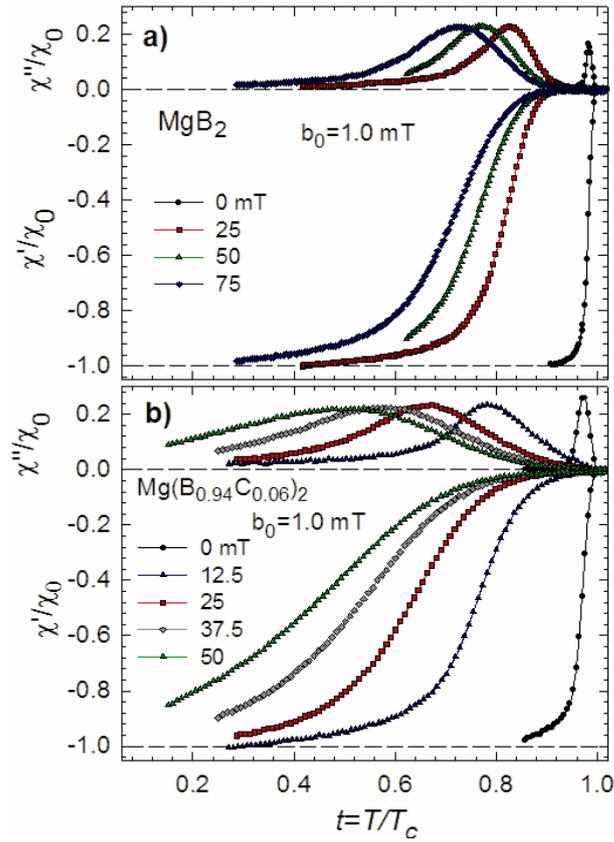

**Fig. 4**. Influence of the DC bias magnetic field on the complex AC magnetic susceptibility for the unsubstituted and C-substituted $MgB_2$ single crystals. The data are obtained for the amplitude of 1.0 mT and frequencies of 96 Hz.

Figure 4 shows the results of the AC susceptibility measurements for the AC magnetic field amplitude $b_0$=1.0 mT, in the presence of DC bias magnetic fields from 12.5 mT to 75 mT. Here, similar to Figure 3, only one, single peak is observed in $\chi''$ vs $t$ dependence, and its width broadens when the DC magnetic field increases. This broadening is much more pronounced for the MBC crystal. For this crystal, the DC bias magnetic field lowers slightly the maximum in $\chi''(t)$, *e.g.* from $\chi''\approx 0.26$ to about 0.22 for field increasing from 0 to 50 mT. In turn, for the MB crystal, we observe opposite behavior; an increase of $\chi''$ from an initial value of 0.17 to 0.223, for the DC bias field of 25 mT. The main observation is that the C-substituted $MgB_2$ is more resistant to the magnetic field, thus at high fields should show superior critical parameters ($J_c$, $H_{irr}$) than the unsubstituted compound, despite of inferior parameters at low fields. This has been confirmed by results obtained for unsubstituted and C-substituted $MgB_2$ at fields higher than about 1-2 T, depending on C content [21-22].

**3.3 Critical current densities derived from AC susceptibilities data**

In general, a point where $\chi''$ shows its maximum on $\chi''(T, b_0)$ characteristics is strictly related with intrinsic properties of a superconductor (mainly pinning force), its temperature and geometry in respect to direction of an applied AC driving magnetic field, the field amplitude and frequency. According to the critical state model, the maximum at $\chi''$ occurs when the magnetic flux front inside a type-II superconductor is just reaching the center of the superconductor of an arbitrary dimension $a$. This situation can be described by a general relation $H_p \propto J_c \cdot a$, where $H_p$ denotes the full penetration field and $J_c$-the critical current density, for $a$ perpendicular to the field direction. For considerably simple and well defined geometries of superconducting samples (*e.g.* long cylinders, plates, thin films and disks, *etc.*) and for the Bean-London critical state model, interrelations between AC complex susceptibility, critical current density, dimensions of the sample and applied magnetic field can be calculated analytically [23]. In a case of more complex configurations numerical calculations are required. For interpretation of our experimental results, we adopted theoretical approach given by Pardo *et al.* [24], for the case of a rectangular bar in an external transverse magnetic field. Figure 5 shows mutual relations between full penetration field, critical current density, transverse dimension of the sample *a* and its thickness *c*, and an amplitude $b_0(\chi_{max}'')$ at which a maximum in $\chi''(T,b_0)$ occurs. Although detailed numerical calculations were performed for a rectangular bar of one side going to infinity, $b\to\infty$, the results can be applied, with a small error, also for samples with dimension $2b>2a$, as in the case of our crystals (see Table 1). Using relations of Fig. 5 and experimental data presented in Figs. 3 and 4, one can derive critical current densities as a function of temperature and external magnetic fields. Such calculated values of $J_c$ vs. temperature are shown in Figure 6. For no DC bias external magnetic field, $J_c(T)$ drops linearly with *T* at temperatures close to the transition point and at low magnetic field amplitudes used in our experiments, *i.e.* from 0.2 to 1.7 mT. This can be approximated by the linear relation $J_c(t) \propto J_c(0)(1-t)$, where $t=T/T_c$. Close to the transition temperature, *i.e.* from $T_c$ up to $t=0.98$ (unsubstituted crystal) and up to $t=0.95$ (carbon substituted crystal), $J_c$ drops with temperature about 1.6 faster for the unsubstituted $MgB_2$ crystal than for the crystal substituted with carbon. For the data without DC bias magnetic field, assuming a linear $J_c(t)$ dependence, the values of $J_c$ extrapolated to 10 and 15 K are $1.0 \times 10^9$ and $8.6 \times 10^8$ A/m$^2$, respectively, for the MB crystal. Similar extrapolation for the MBC crystal gives $J_c$ of $4.7 \times 10^8$ A/m$^2$ at 10 K. As one can see, the values of $J_c$ determined by AC susceptibility and DC magnetic moment measurements are nearly the same (Fig. 2 and Table 1).

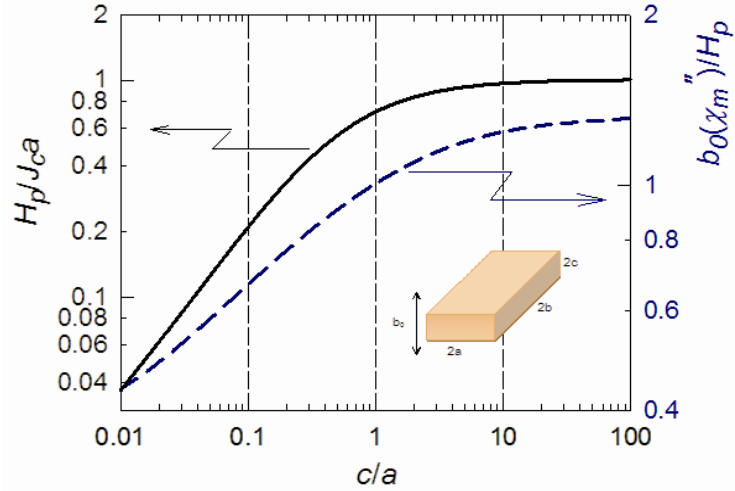

**Fig. 5**. Theoretical relations between $H_p$, $J_c$, dimensions of a sample, and the field where a maximum in $\chi''(T,b_0)$ occurs, $b_0(\chi_m'')$, after Pardo *et al* [24].

Temperature dependencies of the critical current densities in various DC magnetic fields have been determined in a similar way as above from the susceptibility data shown in Figure 4. Application of the external DC bias magnetic field causes a change in the temperature dependence of the critical current density; from a linear dependence at temperatures close to the transition to an exponential character, $J_c(t) \propto J_c(0)(1-t)^p$. The coefficient *p* depends on the DC bias magnetic field and takes the values approximately from 3.4 to 2.4 at fields from 25 to 75 mT (see Fig. 6). For the DC fields used in our experiments, we observe the $J_c(t)$ dependence characteristic for low pinning superconductors; relatively large exponent *p*, which means relatively strong decay of the critical current density with temperature.

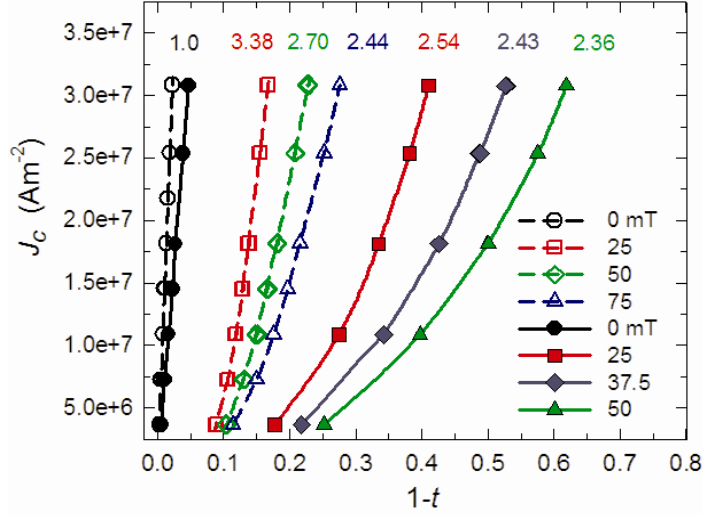

**Fig. 6**. Critical current densities as a function of the reduced temperature, $t=T/T_c$, derived from the peaks in the imaginary part of the AC susceptibility data, $\chi''_{max}$ (see Figs. 3 and 4). Open and filled symbols are for unsubstituted and C-substituted $MgB_2$ single crystals, respectively. The curves are labeled with $p$ values in the relation $J_c \propto (1-t)^p$.

### 3.4 Energy losses

The imaginary part of the AC susceptibility, $\chi''$, is related to the energy losses as:

$$Q_v = (\mu_0)^{-1} \pi \chi_0 \chi'' b_0^2 \qquad (3)$$

where $Q_v$ are losses per unit volume and per cycle, $\mu_0$ is the magnetic constant ($4\pi \cdot 10^{-7}$ H/m). The product $\chi_0 \cdot \chi''$ is just a raw value of the imaginary part of the AC susceptibility as measured by a susceptometer. In turn, $\chi''$ is a quantity related to the intrinsic magnetic properties of the superconductor (internal susceptibility) and tells what part of the external magnetic field energy is dissipated in the superconductor as Joule heat. For type-II superconductors, the values of $\chi''$ range from very low values to about 0.5, depending on the superconducting material and external electromagnetic conditions [20,23,25]. Taking into account data of Figures 3 and 4, and relation (3) $Q_v$ vs. $b_0$ loss characteristics were determined and plotted as in Figure 7.

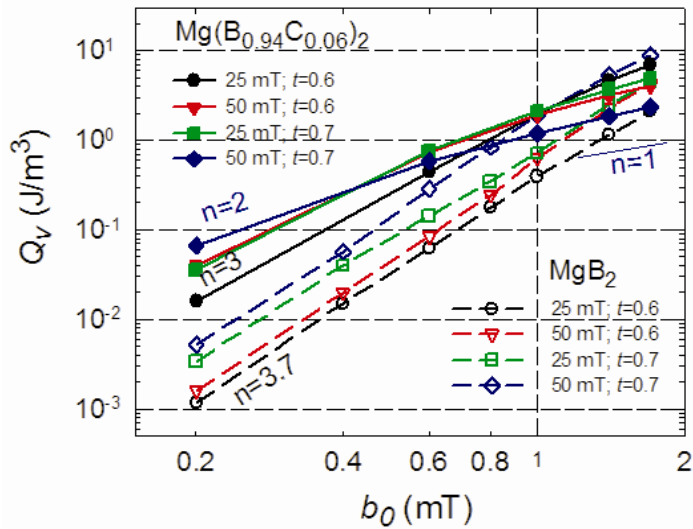

**Fig. 7**. AC volume losses $Q_v$ versus magnetic field amplitude $b_0$, for the $MgB_2$ single crystals, unsubstituted (open symbols and dashed lines) and substituted with carbon (filled symbols and solid lines). The data were taken at two temperatures ($t=T/T_c=0.6$ and 0.7) and two DC bias magnetic fields (25 and 50 mT).

In general, magnetization losses depend on the amplitude of an external magnetic field, $b_0$, in a power form; $Q_v \propto b_0^n$, where coefficient $n$ takes usually values of 1 to 4, in most cases of the losses in type-II superconductors [25]. As shown in Figure 7, the MB crystal is not fully saturated at the applied field amplitudes and at temperatures of 0.6 and 0.7 $T_c$. Here, the $Q_v(b_0)$ curves have nearly the same slope of about 3.7 in a log-log scale plot. One can assume that such power dependence of the losses ($n \cong 4$) is caused by the Kim-like critical state rather, than resulting from geometry of a thin film in perpendicular magnetic field. Average aspect ratios ($a/c$ or $b/c$) of our single crystals are of the order of 5-12 (see Table 1), so considerable lower than for typical thin film samples, *e.g.* coated conductors, where the aspect ratios are higher than 100 [26]. The behavior of the MBC crystal is different. Here, for the given values of the magnetic fields and temperatures, the coefficient $n$ changes from about two (at low $b_0$) to one (at higher $b_0$), so the crystal is just in the transition region, *i.e.* it goes from the partially saturated state to the state of the full penetration (and beyond). The results in Figure 7 show evidently, that for higher magnetic fields, the losses in the MBC crystal are lower than losses in the MB crystal. This observation seems to be important for low AC field and current applications. The values of the losses presented in Figure 7 are in good agreement with corresponding magnetization loss data obtained for sintered, polycrystalline $MgB_2$ samples [27-28].

## 4 Conclusions

Magnetic moment hysteresis loops were measured for the unsubstituted and C-substituted $MgB_2$ single crystals at 10 and 15 K and in fields up to 0.4 T. From these loops the critical currents, $J_c$, have been derived by using the Bean critical state model. We found that the dependence of $J_c$ on the external magnetic field can be described well by the relation of the power form: $J_c(B) \propto (B/B_0)^{-\beta}$, where $B_0$ and $\beta$ are parameters depending on the microstructure properties of a superconductor. For the unsubstituted $MgB_2$ single crystal, the best $J_c(B)$ fits were obtained with parameters $B_0$=43 and 36 mT and $\beta$=2.15 and 2, at the temperatures of 10 and 15 K, respectively. For the C-substituted crystal, the best $J_c(B)$ fits were found for $B_0$=21 mT and $\beta$= 2.6 at the temperature of 10 K. The results show that the substitution of $MgB_2$ with carbon, which lowers considerably the transition temperature, decreases critical current densities at low external magnetic fields. This effect is opposite to that observed at high magnetic fields, where the significantly increased critical current densities have been obtained for the C-substituted $MgB_2$ compound.

Measurements of the AC complex magnetic susceptibility were carried out at the AC magnetic field with amplitude ranged from 0.2 to 1.7 mT and frequency from 24 to 768 Hz, and at the DC bias magnetic field up to 75 mT. At the low frequency range (up to ~100 Hz) and at low AC magnetic field amplitudes (up to ~1.0 mT) the AC susceptibility is frequency independent. Using the results $\chi''(T, b_0)$ we derived critical current densities as a function of temperature and external magnetic field. These critical current densities are very similar to those obtained from magnetic moment hysteresis loops. For zero DC bias magnetic field, $J_c(T)$ drops linearly with $T$ at temperatures close to $T_c$ and at low AC magnetic field amplitudes, for both unsubstituted and C-substituted $MgB_2$ single crystals.

The AC volume losses were determined at temperatures of 0.6 and 0.7 $T_c$, so at temperatures close to the boiling points of $LH_2$ and LNe, potential cooling media for magnesium $MgB_2$ based systems. At higher magnetic fields the losses for the C-substituted crystal are lower than losses for the unsubstituted crystal. However, at low fields the unsubstituted crystal shows lower losses and higher critical currents, thus unsubstituted $MgB_2$ seems to be better material but for low-field/low-current applications only.

This work was supported by the Polish Ministry of Science and Higher Education under research projects for the years 2006-2009 (Project N202 131 31/2223) and, partially, for the years 2009-2011 (MC and KR, Project N N510 357437) and 2009-2010 (KO, project N N202 235937).